\documentstyle[ApJ,Times]{article}

\begin{document}

\def\simg{\mathrel{%
      \rlap{\raise 0.511ex \hbox{$>$}}{\lower 0.511ex \hbox{$\sim$}}}}
\def\siml{\mathrel{%
      \rlap{\raise 0.511ex \hbox{$<$}}{\lower 0.511ex \hbox{$\sim$}}}}
\def\Mesz{M\'esz\'aros~} \def\hsp{\hspace*{2mm}}
\def\ie{i.e$.$~} \def\eg{e.g$.$~} \def\etal{et al$.$~} \def\eq{eq$.$~} 
\def\Lunits{\rm{erg\,s^{-1}}~} \def\Cunits{\gamma\,{\rm cm^{-2} s^{-1}}~}

\title{ Power Density Spectra of Gamma-Ray Bursts in the Internal Shock Model }

\author{A. Panaitescu\altaffilmark{1}, M. Spada\altaffilmark{1,2} \& 
                    P. \Mesz\altaffilmark{1}}
\affil{Department of Astronomy \& Astrophysics, Pennsylvania State University, 
          University Park, PA 16802}
\altaffiltext{1}{also Institute of Theoretical Physics, University of California, 
          Santa Barbara}
\altaffiltext{2}{also Osservatorio di Arcetri, Universit\'a di Firenze, Italy}

\begin{abstract}

 We simulate Gamma-Ray Bursts arising from internal shocks in relativistic winds,
calculate their power density spectrum (PDS), and identify the factors to which the 
PDS is most sensitive: the wind ejection features, which determine the wind dynamics 
and its optical thickness, and the energy release parameters, which give the pulse 
50--300 keV radiative efficiency. For certain combinations of ejection features and 
wind parameters the resulting PDS exhibits the features observed in real bursts. 
We found that the upper limit on the efficiency of conversion of wind kinetic energy 
into 50--300 keV photons is $\sim$ 1\%. Winds with a modulated Lorentz factor distribution 
of the ejecta yield PDSs in accord with current observations and have efficiencies 
closer to $10^{-3}$, while winds with a random, uniform Lorentz factor ejection must 
be optically thick to the short duration pulses to produce correct PDSs, and have an 
overall efficiency around $10^{-4}$. 

\end{abstract}

\keywords{gamma-rays: bursts - methods: numerical - radiation mechanisms: nonthermal}

\section{Introduction}

 Internal shocks occurring in a transient, unstable relativistic wind (Rees \& \Mesz 1994) 
are believed to be the source of the $\sim 100$ keV emission and the complex temporal 
structure of Gamma-Ray Bursts (GRBs). The shocks resulting from such instabilities heat the 
expanding ejecta, amplify pre-existing magnetic fields or generate a turbulent one, and 
accelerate electrons, leading to synchrotron emission and inverse Compton scatterings. 
Synchrotron self absorption and pair formation may further shape the emergent spectrum, 
depending on the choice of model parameters, as shown by Papathanassiou \& \Mesz (1996)
and Pilla \& Loeb (1998). The spectrum of the emitted radiation has also been analyzed 
by Daigne \& Mochkovitch (1998) and Panaitescu \& \Mesz (1999). The synchrotron 
self-Comptonized emission from internal shock GRBs has also been studied by Sari \& Piran
(1997) and Ghisellini \& Celotti (1999). 

 In this paper we analyze the temporal features of GRBs obtained in the framework of 
relativistic winds, through the burst power density spectrum (PDS). The recent work of 
Beloborodov, Stern \& Svensson (1998) has put into evidence interesting PDS 
features, which we use for analyzing some of the model parameters and the features of the 
wind ejection.  We use the observed integral burst intensity ($\log N -\log P$) distribution 
(Pendleton \etal 1996) as a constraint in our choice for some of the model parameters. 
Other properties of the observed light curves, such as the distribution of pulse durations
(Norris \etal 1996) or those of pulse fluences and time intervals between peaks 
(Li \& Fenimore 1996), can also be used in the study of the wind ejection features
(Spada, Panaitescu \& \Mesz 1999).

\section{Outline of the Model Features}

 The model used here is similar to that developed by Daigne \& Mochkovitch (1998), and is 
different in the following ways. Our treatment of the radiation emission takes into account 
the up-scattering of the synchrotron photons, which may be the dominant emission process at 
$\sim 100$ keV if the magnetic field is sufficiently low. The optical thickness (to Thomson 
scatterings) of the wind is taken into account, which may be very important for the burst PDS 
and overall efficiency. We use the shock jump equations to determine the physical conditions 
in the shocked fluid and we include the effects arising from electron cooling in the shape 
of the emission spectrum.

 After setting the dynamics of the wind ejection, we calculate the radii where internal 
collisions take place, determine the relevant physical parameters in the shock fluid -- 
bulk Lorentz factor (LF), typical electron random, magnetic field, etc$.$ -- and calculate 
the features of the emitted radiation: energy of the synchrotron and inverse Compton spectral 
peaks (with allowance for the Klein-Nishina effect), Comptonization parameter, 50--300 keV
emission (accounting for scattering during propagation through the wind), and its duration. 
These quantities are necessary for the computation of the observer frame pulse features:
fluence, duration, arrival time, taking into account relativistic and cosmological effects.
Assuming the two-sided exponential pulse shape identified by Norris \etal (1996) in real
bursts, one can calculate the peak photon flux for each pulse. Below we describe the most 
important aspects related to the calculation of the some of the above mentioned quantities.

 The wind is discretized as a sequence of shells, released by the central source during a time 
$t_w$ and with an average interval $t_v$ between consecutive ejections, so the number of shells 
is just $N=t_w/t_v$. The mass of each shell is drawn from a log-normal distribution determined 
by the average shell mass $\overline{M}=M_w/N$, $M_w$ being the wind mass, and a dispersion 
$\sigma_M= \overline{M}$. The LF $\eta$ of each shell is randomly drawn from the interval 
$[\eta_m,\eta_M]$. Both $\eta_m$ and $\eta_M$ can be chosen constant during the entire wind 
ejection, so that $\eta$ has a uniform distribution (we shall refer to this case as the "uniform 
wind"), or can vary on time-scale comparable to $t_w$ and much larger than $t_v$, according to 
a certain law. For brevity, in the latter case, we shall keep $\eta_m$ constant and 
use an $\eta_M$ that varies with ejection time as a square sine with only one oscillation (this 
will be the "modulated wind"), thus the LF of the $j^{\rm th}$ ejected shell is
\begin{equation}
 \eta_j = \eta_m + a_j \sin^2 (\pi j/N) (\eta_M-\eta_m) \quad 1 \leq j \leq N \;,
\label{sinesq}
\end{equation}
where $a_j$ is a random number between 0 and 1.
The time elapsed before the ejection of a shell is chosen to be proportional to the energy 
of that shell, thus leading to a wind luminosity $L_w$ that is constant on average throughout the 
entire wind ejection.

 The shock-accelerated electrons have a power-law distribution of index $-p$, starting from 
a low random LF set by the electron injection fraction $\zeta \siml 1$ and the fraction 
$\varepsilon_e$ of the internal energy stored in the electrons:
\begin{equation}
 \gamma_m = \frac{p-2}{p-1} \frac{\varepsilon_e u'}{\zeta n'_e m_ec^2}
            \stackrel{\epsilon_{dis} \ll 1} {=} 
          1837 \frac{p-2}{p-1} \frac{\varepsilon_e}{\zeta} \epsilon_{dis} \;,
\label{gmmin}
\end{equation}
where $u'$ and $n'_e$ are the comoving frame internal energy density and electron number 
density of the shocked fluid, respectively, which we calculate with the aid of the shock 
jump conditions, and $\epsilon_{dis}$ is the dissipation efficiency, \ie the fraction
of the kinetic energy that is converted into internal. 
The magnetic field is assumed turbulent and parameterized through the fraction 
$\varepsilon_B$ of the internal energy density of the shocked gas stored by it:
\begin{equation}
  B^2 = 8\pi \varepsilon_B u' \stackrel{\epsilon_{dis} \ll 1}{=} 
        8\pi \varepsilon_B \epsilon_{dis} n'_e m_p c^2 \;.
\label{Bmag}
\end{equation}

 For the computation of the burst emission in the 50--300 keV band, we approximate the 
synchrotron spectrum of each pulse as a three power-law function, with slopes that depend 
on $p$ and on the relative values of the cooling and peak frequencies. A fraction 
$\min (1,\tau_{\zeta})$ of the synchrotron photons undergoes $\max (1,\tau_{\zeta}^2)$ 
up-scatterings in which their energy is increased by a factor $\sim (4/3) \gamma_m^2$ per 
scattering (unless the Klein-Nishina regime is reached), where $\tau_{\zeta} = \sigma_{Th} 
(\zeta n'_e) \min (c t'_{\gamma}, \Delta')$ is the optical thickness to hot electrons, 
$\Delta'$ being the co-moving shell thickness, $t'_{\gamma} = t'_{sy}/(1+y)$ the radiative 
cooling time-scale, $t'_{sy} \propto (\gamma_m B^2)^{-1}$ the synchrotron cooling time-scale 
and $y$ the Comptonization parameter, which depends on $\tau_{\zeta}$ (for $\tau_{\zeta} < 1$, 
$y=\gamma_m^2 \tau_{\zeta}$). The $\Delta'$ is calculated from the shock compression factor
and from the thickness of the shell before the collision. For the latter we assume that
during the free expansion between successive collisions, the shell thickness fractional 
increase is equal to the fractional increase of its radius $r$ 
(${\rm d}\log \Delta'/{\rm d}\log r=1$).

 We take into account the optical thickness to Thomson scatterings on cold electrons in 
the emitting shell ($\tau_0$) and in the outer part of the wind through which a pulse 
propagates ($\tau$), by diminishing the pulse intensity by a factor $(1-e^{-\tau_0}) 
\tau_0^{-1}$ and $e^{-\tau}$, respectively. However we do not include in the burst 
light-curve calculation the scattered photons which, in the end, still arrive at the 
observer. If the cumulative time delay due to scattering/diffusion through the wind is 
larger than the burst duration then, for the observer, the photon scattering practically 
acts like an absorption process. Otherwise it may mimic an emission process with a timescale 
shorter or comparable to the burst duration, and one should take it into account in 
the light-curve (and PDS) calculation.

 Very important for the burst PDS is the computation of the pulse duration,
which is determined by: \\
\hsp 1) the geometrical curvature of the emitting shell (the observer receives most of the
    radiation from a zone extending up to an angle $\sim \Gamma^{-1}$ relative to the central 
    line of sight), \\
\hsp 2) the lab-frame electron radiative cooling time $t_{\gamma} = \Gamma t'_{\gamma}$, \\
\hsp 3) the lab-frame shock's shell-crossing time $t_{\Delta}= \Delta /|v_{sh}-v_0|$, $v_{sh}$ 
    being the shock's speed, determined from the hydrodynamics of the collision, and $v_0$ the 
    shell pre-shock flow velocity. \\
The spreads in the photon arrival time are functions of the angle relative to the central line 
of sight; so is the intensity of the relativistically boosted emission. The intensity-averaged 
observer time spreads corresponding to the three factors above are
\begin{equation}
 \delta T_{\theta} \sim \frac{r}{2\Gamma^2 c}, \quad
 \delta T_{\gamma} \sim \frac{t_{\gamma}}{\Gamma^2}, \quad 
 \delta T_{\Delta} \sim \frac{t_{\Delta}}{\Gamma^2} \;,
\label{deltaT}
\end{equation}
which we add in quadrature to calculate the pulse duration $\delta T$. 
For further qualitative estimations we will consider that the pulse duration is positively
correlated with the radius where the collision takes place. This would be obvious if $\delta T
\simeq \delta T_{\theta}$, but it is a correct assumption because both $\delta T_{\gamma}$ 
and $\delta T_{\Delta}$ have the trend of increasing with $r$, due to the fact that 
the electron LFs/magnetic fields decrease with radius, while the shell thickness increases.

 The 30--500 keV pulse fluence is a fraction of the kinetic energy of the colliding shells,
equal to the product of \\
\hsp 1) the {\sl dissipation efficiency} $\epsilon_{dis}$. If $\eta_M \gg \eta_m$ then 
    $\epsilon_{dis}$ can exceed 10-20\% for the collisions taking place at small radii, where 
    there is a larger difference between the shell LFs, decreasing to 1\% or less for the late 
    collisions. \\
\hsp 2) the {\sl radiative efficiency} at which the internal energy is converted into radiation.
    It is upper bounded by $\varepsilon_e$, the fractional energy in electrons, reached
    if the $\gamma_m$-electrons are radiative. \\
\hsp 3) the {\sl window efficiency} $\epsilon_{2+3}$, representing the fraction of the radiated 
    energy that arrives at observer in the 50--300 keV band (the ${\rm 2^{nd}}$ and ${\rm 3^{rd}}$ 
    BATSE channels). Given the broadness of the synchrotron and inverse Compton spectra,  
     $\epsilon_{2+3}$ cannot exceed 30\% for $p=2.5$. \\
Thus the overall burst efficiency can hardly exceed 1\%, a value reached if the model 
parameters are such that for the most energetic collisions in the wind the above 
efficiencies are close to their maximal values.

\section{Numerical Results}

 Beloborodov \etal (1998) (hereafter BSS98) have calculated the average PDS of 214 GRBs longer 
than 20 seconds using their 50--300 keV light-curves with 64 ms resolution. The PDS they 
obtained follows a power-law $P_f \propto f^{-5/3}$ ($f$ is frequency) from $f_l \sim 0.03$ Hz 
to $f_h \sim 1\,{\rm Hz}$, decaying weaker below $f_l$ and stronger above $f_h$. In
other words, $f^{5/3} P_f$ is flat up to $f_h$, where a  break is observed. We have 
identified three factors that can lead to such PDSs: $i)$ the pulse window efficiency 
$\epsilon_{2+3}$, $ii)$ the LF ejection law, which determines the wind dynamics and the 
dependence of $\epsilon_{dis}$ on $\delta T$, and $iii)$ the wind optical thickness $\tau$ 
for short pulses.

 Figure 1$a$ shows the effect of $\zeta$ and $\varepsilon_B$ on $\epsilon_{pulse}$, 
the fraction of the wind entire kinetic energy that is radiated by each pulse in the 50--300
keV band. The $\epsilon_{dis}$ and $n'_e$ decrease with $r$, thus $\gamma_m$ (\eq [\ref{gmmin}]) 
and $B$ (\eq [\ref{Bmag}]) exhibit the same trend, which makes $\epsilon_{2+3}$ decrease 
with $\delta T$. The two energy release parameters $\zeta$ and $\varepsilon_B$ alter the peak 
energies of the synchrotron and inverse Compton spectra (they are higher for lower $\zeta$) 
and the Comptonization parameter (which is larger for smaller $\varepsilon_B$). In the case 
shown here, decreasing these parameters results in increasing the $\epsilon_{2+3}$ of the 
longer pulses. The PDSs in Figure 1$b$ illustrate how the burst power is 
distributed versus variability timescale, by combining the pulse duration distribution
with the $\epsilon_{pulse}$--$\delta T$ dependence of Figure 1$a$. The conclusion to be
drawn from Figure 1$b$ is that, for uniform winds that are optically thin 
to most pulses, electron fractions below unity and magnetic fields well below equipartition 
reduce the power in the short pulses, thus representing one possible cause for the 
1 Hz break; nevertheless the PDS remains flat below 1 Hz.

 An ejection with modulated LFs mitigates the decrease of $\epsilon_{dis}$ with radius, 
by clumping shells earlier in the wind expansion and producing groups of shocked shells that 
travel longer distances before suffering more collisions. Thus the differences between the LFs
of the shells colliding at larger radii are greater, yielding a better $\epsilon_{pulse}$ for 
longer pulses. This is illustrated in Figures 1$c$ and 1$d$ for winds that are optically thin, 
where we used a LF ejection modulated by a sine square (\eq [\ref{sinesq}]). Note the 
flatness of $f^{5/3} P_f$ at $f \siml 0.5$ Hz.

 The fact that $\delta T$ and the radius where the collision takes place are positively 
correlated suggests that the observed lack of power in short duration pulses may also be due to 
the large $\tau$ of the wind at early lab-frame times. We shall refer to the case where short 
pulses occur predominantly below the photospheric radius as the "optically thick case", without
 meaning that the wind is thick for all pulses.  Equation (3) from Rees \& \Mesz (1994) gives 
an estimation of the photospheric radius $r_{ph} = (\kappa L_w)/(8\pi c^3 \overline{\eta}^3)$, 
where $\kappa=0.4\,{\rm cm^2 g^{-1}}$ and $\overline{\eta}$ is the wind average LF. If $\delta T$ 
is mainly determined by the geometrical curvature of the emitting shell, then the duration of the 
pulses emitted at the photospheric radius is $\delta T_{ph} \sim r_{ph}/(2\overline{\eta}^2 c)$ :
\begin{equation}
 \delta T_{ph} = \frac{\kappa L_w}{16\pi c^4 \overline{\eta}^5}
               = 1.0\,L_{w,54} \overline{\eta}_2^{-5} \, {\rm s} \;. 
\label{dTph}
\end{equation}
Equation (\ref{dTph}) shows that the duration of the pulses for which the wind is optically 
thick is very sensitive to $\overline{\eta}$. An increase in the wind's $\tau$ may 
result from a higher $L_w$ or a lower $\overline{\eta}$, either case implying a higher wind 
mass. Numerically we found that for an uniform wind with $\eta_m=30$ and $L_w = 10^{54}
\,\Lunits$ $\eta_M$ must be below 300 to increase the wind's $\tau$ to the point where 
there is substantial loss of power at the high frequency end of the PDS. Figures 1$e$ and 1$f$ 
show that this loss of power can be enhanced if the magnetic field is far from equipartition. 

 To account for the effects arising from the cosmological distribution of GRBs in the
calculation of the PDS and intensity distribution, the GRB redshifts are chosen from the 
probability distribution
\begin{equation}
 \frac{dp}{dz} \propto \frac{n_c(z)}{1+z} \frac{dV}{dz} \;,
\label{dpdz}
\end{equation}
where $dV/dz$ is the cosmological comoving volume per unit redshift, and $n_c(z)$ is the rate 
density evolution of GRBs. For brevity, we consider here the case of a constant rate density, 
keeping in mind that other reasonable functions $n_c(z)$ alter the PDS, though the changes are 
not drastic. We use an un-evolving power-law distribution for the wind luminosity:
\begin{equation}
 \Phi(L_w) \propto L_w^{-\beta}\;, \quad L_m \leq L_w \leq L_M \;,
\label{phiL}
\end{equation}
and zero otherwise. Note that this not the same as assuming that the GRB 50--300 keV luminosity
at the source redshift has a power-law distribution, as it is usually done (\eg Reichart \& 
\Mesz 1997, Krumholz \etal 1998).

 Figure 2 illustrates the effect of $L_w$ and $z$ on the PDS, for a $\sin^2$-modulated wind. 
As can be seen, there is a shift of power from higher frequencies to lower ones if $L_w$ is 
increased. The same is true for a burst placed at a higher redshift. The latter is due to the 
time dilation, which makes pulses appear longer. An increase in $L_w$ leaves unaltered the 
wind dynamics and enhances the magnetic field (the comoving particle density in \eq [\ref{Bmag}] 
is higher). Thus the emission becomes harder and the window efficiency will favor longer pulses. 
If the wind is quasi-optically thick, the correlation between the PDS and $L_w$ is 
strengthened by the increase of $\tau$ with $L_w$. 

 Figure 3$a$ shows the intensity distribution of simulated bursts, using the peak fluxes on 
256 ms timescale, compared to the observed distribution (Pendleton \etal 1996). About 50\% of 
the 400 simulated bursts are brighter than $0.4\,\Cunits$, thus the number of bursts in our 
sample is close to that of the bursts analyzed by Pendleton (1996).
Excluding the two points with the lowest peak flux and the one with the highest (which
contains only 1--3 bursts), the $\chi^2$ for the sets of optically thin, modulated and thick,
uniform winds are 7.8 and 16.4, respectively, for 8 degrees of freedom. Thus only the intensity 
distribution of the modulated wind bursts is consistent with the observations. A lower $L_m$ 
would yield a better fit of the $\log N - \log P$ distribution for uniform winds, but the PDS 
would have too much power at high $f$.

 The PDSs shown in Figure 3$b$ have been calculated by averaging the PDS of all bursts with 
peak photon fluxes brighter than $1\,\Cunits$ on the 64 ms timescale. More than one of the 
mechanisms presented in Figure 1 must be invoked in order to obtain $f^{-5/3}$ behavior of 
the PDS and the 1 Hz break: for modulated bursts we used $\zeta=10^{-2}$ and 
$\varepsilon_B=10^{-6}$; for uniform winds $\eta_M=220$ and $\varepsilon_B=10^{-4}$ reduce 
the brightness of the short pulses to the point where $f^{5/3}P_f$ exhibits a flat part, albeit 
over a range of frequencies narrower than observed. In the former case the wind efficiency is 
slightly larger than 0.1\%; in the latter, the scatterings of the photons in the wind diminishes 
the burst efficiency to $\sim 10^{-4}$. For this reason, thick un-modulated winds require a 
larger energy budget than the thin modulated ones.

 Figure 3$c$ shows how the power spectra of individual bursts are spread around the average 
PDS. The distribution was calculated by averaging the $P_f$ distributions obtained for a large
number of frequencies. This figure also shows the exponential 
fit $dN/d\log P_f = (\ln 10) (P_f/\overline{P}_f) \exp(-P_f/\overline{P}_f)$ that BSS98 found 
to approximate well the observed distribution. Note that the two models for ejection LF partly 
bracket the exponential fit, suggesting that in real bursts the wind is neither as modulated 
as the squared sine employed here, nor as erratic as the uniform distribution, but somewhere 
in between. This conclusion is consistent with the fact that the observed 1 Hz break is in 
between the breaks exhibited by the simulated PDSs shown in Figure 3$b$.

 In Figure 4 we show the typical 50--300 keV light-curves of bursts with a modulated and an
uniform wind.

\section{Conclusions}

 We have investigated the power spectra of the $\gamma$-ray light curves expected from 
internal shock GRBs. The PDS is found to be most sensitive to the wind luminosity $L_w$,
its average LF $\overline{\eta}(\eta_m,\eta_M)$, and the burst redshift $z$.  The PDS is 
sensitive to other model parameters, such as the wind duration $t_w$, if the wind LF is 
modulated, and the variability timescale $t_v$, if the LF distribution is just random. 
The properties of the PDS are also affected by the dynamics of the unsteady relativistic 
wind and by the microscopic parameters of the shocked fluid.

 We have compared our results to those of Beloborodov \etal (1998), and have identified 
three possible reasons for the observed deficit of short ($\delta T \siml 1$ s) pulses: \\
\hsp 1) the 50--300 keV radiative efficiency of such pulses may be smaller than for longer
    pulses, due to low electron injection fractions and magnetic fields well below 
    equipartition. \\
\hsp 2) the short pulses may result from the collision of light shells, carrying little 
    energy, due to a modulation of the ejection LFs. Winds with modulated ejection LFs 
    (here we used a squared sine) yield PDSs and $\log N - \log P$ distributions consistent 
    with the observations, and have a 50--300 keV efficiency $\simg 10^{-3}$. \\
\hsp 3) the short pulses may occur below the photospheric radius.  The duration of pulses 
    occurring at the photospheric radius depends strongly on the average LF of the wind 
    (\eq [\ref{dTph}]). For a wind LF distribution that has a lower bound $\eta_m=30$, an 
    upper limit $\eta_M = 220$ leads to a significant decrease of the burst power at high 
    frequency, yielding a PDS behaving like $f^{-5/3}$ for 0.2 Hz $ < f < $ 2 Hz, and to a 
    50--300 keV efficiency around $10^{-4}$. A lower $\eta_M$ yields a PDS with even lower 
    power at high frequency, but the wind efficiency in the middle BATSE channels becomes
    too small. 

 The overall burst efficiency is small, due to the low dissipation efficiency of the wind, 
typically between 1\% and 10\%, and to the broadness of the synchrotron self-Comptonized 
spectra from power-law distributions of electrons, leading to window efficiencies $\sim$ 10\%. 
For the wind luminosities and durations used in the calculations whose results are shown in 
Figure 3, a beaming factor ranging from $\sim 10$ (for bursts with $L_w=L_m$) to $\simg 
1000$ (for $L_w=L_M$) is needed to maintain the energy requirements below the upper 
limits found by \Mesz, Rees \& Wijers (1999) for the energy that can be extracted 
from plausible GRB progenitors. 

 The study of the power spectra of internal shock GRBs has shown that if the ejection
parameters of optically thin winds are totally random, the resulting spectrum is flat,
with equal power at low and high frequency. In order to explain the observed 
$f^{-5/3}$ behavior of the PDS, the wind must be modulated such that collisions at
large radii release more energy in the observing band than the collisions occuring as 
small radii. A modulation of the wind ejection is physically quite plausible, and
the fact that it is necessary to introduce this in order to obtain the correct PDS 
is one indication of the value of the power spectrum as a tool in studying the physics 
of the GRB ``central engine". 

\acknowledgements{This research is supported by NASA NAG5-2857, NSF PHY94-07194
and the CNR. We are grateful to Martin Rees and Steinn Sigurdsson for stimulating 
comments.}

\input PSfig

\begin{figure*}
\vspace*{1cm} 
\centerline{\psfig{figure=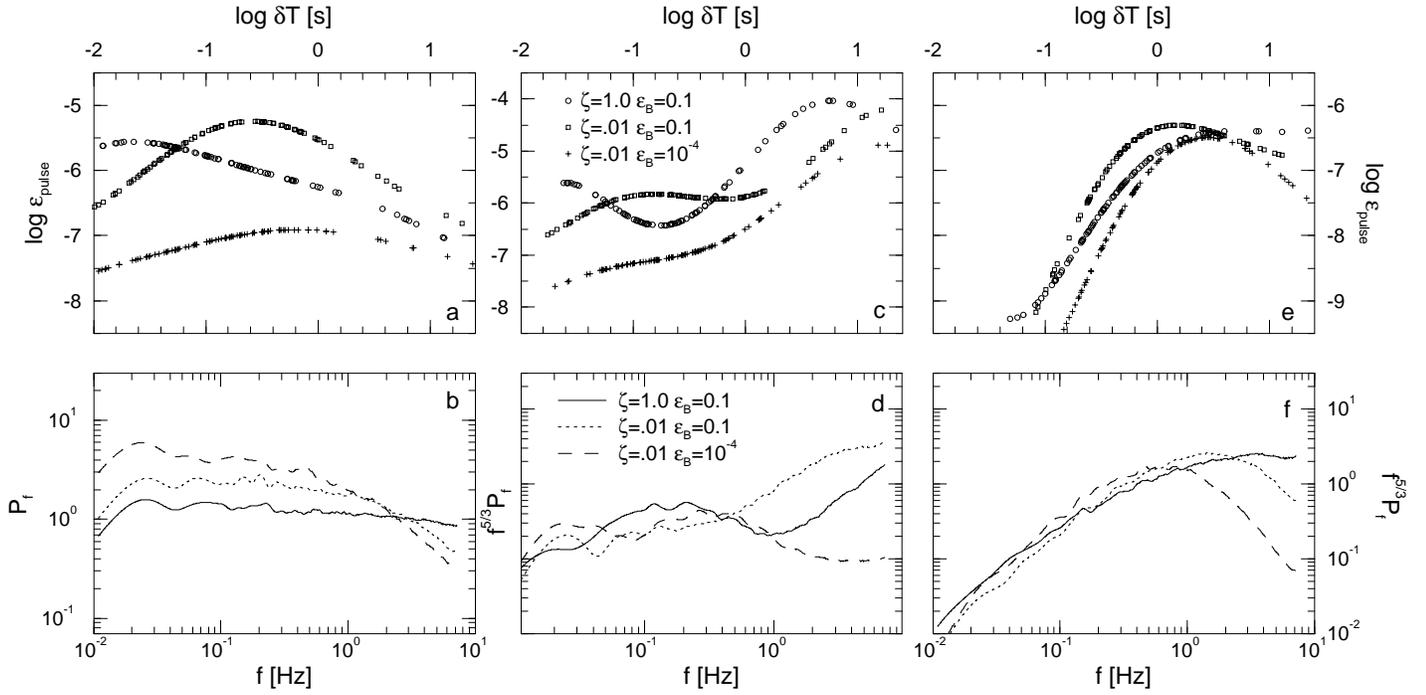}}
\caption{{\sl Upper panels}: the fraction of the wind energy radiated by a pulse in the
        50--300 keV band as function of pulse duration. Parameters: redshift $z=1$,
        $\eta_m=30$, $\varepsilon_e=0.25$, $p=2.5$, $t_w=20$ s, $t_v=20$ ms ($N=10^3$
        shells).  Only a small fraction of the total number of pulses is shown; the density 
        of the points illustrates the pulse duration distribution. The curves shown are 
        log-log space fits for the most efficient pulses. The actual values are scattered 
        around the fit.
       {\sl Lower panels}: the corresponding PDS (note that $f^{5/3}P_f$ is plotted in panels
        1$d$ and 1$f$), calculated by averaging the PDSs of 100 bursts in each case, all
        bursts having the same parameters. Panels 1$a$ and 1$b$: optically thin uniform winds, 
        panels 1$c$ and 1$d$: optically thin modulated winds; for both cases $\eta_M=1000$ and
        $L_w=10^{53}\,\Lunits$. Panels 1$e$ and 1$f$: quasi-optically thick uniform winds
        ($\eta_M=300$, $L_w=10^{54}\, \Lunits$). Other parameters are given in the legend.}
\end{figure*}

\begin{figure*}
\centerline{\psfig{figure=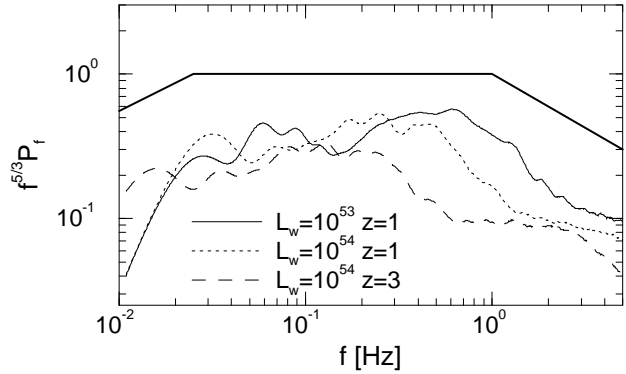}}
\caption{Effect of $L_w$ and $z$ on the PDS of optically thin modulated winds. The solid thick 
       line shows the features (\ie only trends, the ordinate values depend on the choice of 
       normalization) of the PDS obtained by BSS98 for real bursts.
       Other parameters: $\zeta=10^{-2}$, $\varepsilon_e=0.25$, $p=2.5$, 
       $\varepsilon_B=10^{-6}$, $\eta_m=30$, $\eta_M=1000$, $t_w=20$ s, $t_v=20$ ms. 
       Each PDS is the average of 250 power spectra of simulated bursts.}
\end{figure*}

\begin{figure*}
\vspace*{2cm} 
\centerline{\psfig{figure=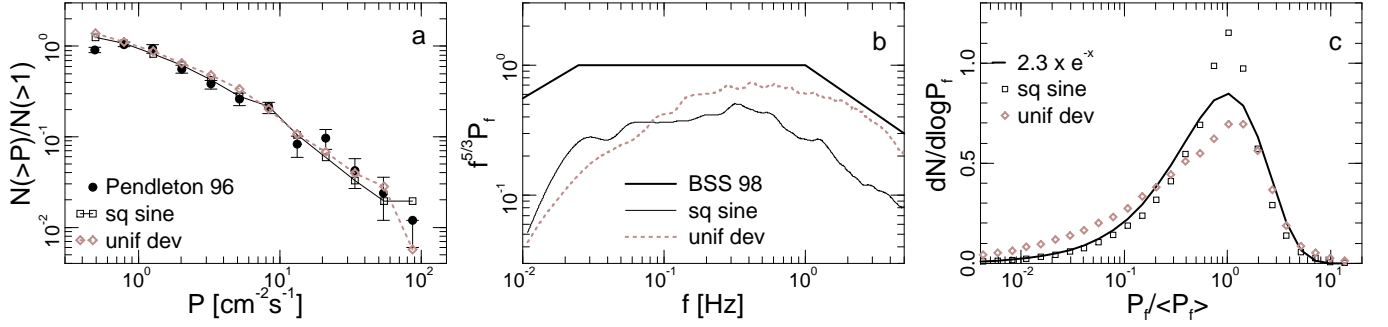}}
\caption{Intensity distribution (panel $a$), average power density spectrum (panel $b$) 
        and distribution of $P_f$ around the average PDS (panel $c$) for two models: 
        $\sin^2$-modulated wind (squares or solid curve) and uniform wind (diamonds or dotted 
        curve). The winds have a power-law luminosity distribution and an un-evolving rate density. 
        The mean redshift and dispersion of the bursts 
        brighter than $1\,\Cunits$ are $z=1.24$, $\sigma_z=0.91$ for the modulated wind model
        and  $z=0.94$, $\sigma_z=0.84$ for the uniform wind model. The former model has 
        $L_m=7\times 10^{52}\,\Lunits$, $\eta_M=1000$, $\varepsilon_B=10^{-6}$, $\zeta=10^{-2}$, 
        the latter is characterized by $L_m=8\times 10^{53}\,\Lunits$, $\eta_M=220$, 
        $\varepsilon_B=10^{-4}$, $\zeta=1$. Other parameters are $L_M/L_m=100$, $\beta=2$, 
        $\eta_m=30$, $t_w=20$ s, $t_v=20$ ms, $\varepsilon_e=0.25$, $p=2.5$, for both models.} 
\end{figure*}

\begin{figure*}
\vspace*{-1cm} 
\centerline{\psfig{figure=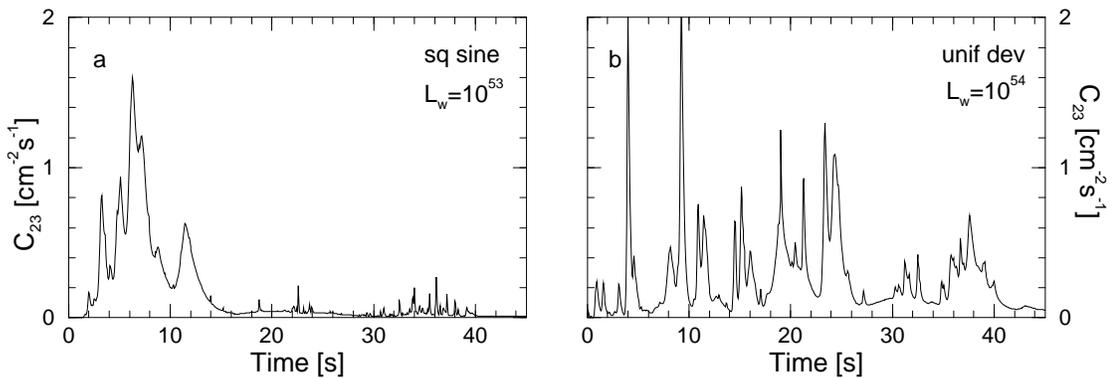}}
\caption{Typical 50--300 keV light-curves of the bursts whose PDSs are shown in Figure 3,
         within the two models for ejection LFs: panel $a$ -- modulated wind with 
         $L_w=10^{53}\,\Lunits$, panel $b$ -- uniform wind with  $L_w=10^{54}\,\Lunits$.
         Both bursts are placed at redshift $z=1$. Other parameters are as for Figure 3.}
\end{figure*}

\end{document}